# Novel Direct Algorithm for Computing Simultaneous All-levels Reliability of Multistate Flow Networks

Wei-Chang Yeh
Department of Industrial Engineering and Engineering Management
National Tsing Hua University
P.O. Box 24-60, Hsinchu, Taiwan 300, R.O.C.
yeh@ieee.org

*Abstract*—Different types of networks, such as, Internet of Things, social networks, wireless sensor networks, transportation networks, and 4g/5G serve to benefit and help our daily lives. The multistate flow network (MFN) is used to model the network structures and applications. The level $d$ reliability, $R_d$, of the MFN is the success probability of sending at least $d$ units of integer flow from the source node to the sink node and is a popular index for designing, managing, controlling, and evaluating MFNs. The traditional indirect algorithms must have all $d$-MPs (special connected vectors) or ($d$-1)-MCs (special disconnected vectors) first, and then use inclusion-exclusion technique (IET) or sum-of-disjoint product (SDP) in terms of found $d$-MPs or ($d$-1)-MCs to calculate $R_d$. The above four procedures are all NP-Hard and #P-Hard and cannot calculate $R_d$ for all $d$ simultaneously. Thus, in this study a novel algorithm based on the binary-addition-tree algorithm is proposed to calculate the $R_d$ directly for all $d$ simultaneously, eliminating the need of using the above four procedures. The time complexity and demonstration of the proposed algorithm were analyzed with suitable examples. Furthermore, an experiment was conducted on 12 benchmark networks to validate the proposed algorithm.

## I. INTRODUCTION

A multistate flow network (MFN) is a special graph with multistate components representing different performance levels with occurrent probabilities. Various practical systems can be modeled using MFNs, for example, the Internet of Things [1, 2], grid and cloud computing [3], wireless sensor networks [4, 5, 6], transportation systems [7, 8], oil/gas production systems [9], power transmission and distribution systems [10, 11], rework networks [12], and Data Mining [13, 14]. Variations in modern network applications are witnessing a rapid growth and being employed extensively [15, 16, 17]. Hence, MFNs play a vital role in many aspects of day-to-day life [18, 19].

Let $G(V, E, \mathbf{D})$ be an MFN, where $V = \{1, 2, \ldots, n\}$ be the complete vertex set and each vertex is perfect. $E = \{a_1, a_2, \ldots, a_m\}$ is the complete edge set, and each edge may fail based on a predefined probability. Further, **D** is the (edge) state distribution defined by the state level and the related state probability of each edge, while vertices 1 and $n$ denote the source and sink nodes, respectively.

For example, Fig. 1 shows an MFN with $V = \{1, 2, 3, 4\}$, $E = \{a_1, a_2, a_3, a_4, a_5\}$, node 1 as the source node, node 4 as the sink node, and the state distribution **D** as listed in Table 1.

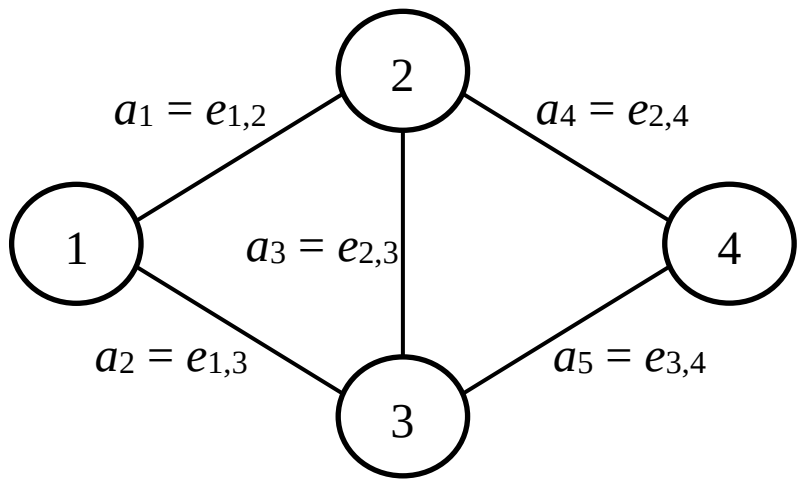

**Figure 1**. The bridge network.

**TABLE 1.** STATE DISTRIBUTIONS IN FIG. 1.

| state | $a_1 = e_{1,2}$ | $a_2 = e_{1,3}$ | $a_3 = e_{2,3}$ | $a_4 = e_{2,4}$ | $a_5 = e_{3,4}$ |
|---|---|---|---|---|---|
| 0 | 0.1 | 0.05 | 0.01 | 0.10 | 0.025 |
| 1 | 0.2 | 0.10 | 0.19 | 0.15 | 0.075 |
| 2 | 0.7 | 0.85 | 0.80 | 0.75 | 0.900 |

Reliability is an extensively used index for evaluating the performance of real-world MFNs [18, 19, 20]. The traditional reliability, $R_d$, of MFNs is defined as the success probability that at least $d$ units of the amount of flows can be sent from vertices 1 to $n$. Without loss of generality, the above problem is referred to as the single-level or level $d$ reliability of MFN. For $d = 1$, this problem can be simply considered as a binary-state network reliability problem.

Various algorithms have been proposed to calculate the single-level reliability of the MFN. It can be calculated by implementing numerous approaches, including exact reliability algorithms [21-35] and approximated reliability algorithms [36, 37, 38]. Most current exact reliability algorithms for calculating $R_d$ are adapted from graph theory and are indirect, that is, either based on cut/($d$-1)-MC [21, 22, 23, 24, 25, 26] or based on path/$d$-MP [27, 28, 29, 30, 31, 32, 33, 34, 35] and thus do not yield $R_d$ directly.

However, in MFNs, it is HP-Hard and #P-Hard to obtain all $d$-MPs or the ($d$-1)-MCs [18, 20], that is, calculating $R_d$ and ($1-R_d$) based on $d$-MPs and ($d$-1)-MCs, respectively, is a challenge. Moreover, after obtaining all $d$-MPs or ($d$-1)-MCs, all current algorithms still need either the inclusion-exclusion technology (IET) [39] or the sum-of-disjoint product algorithm (SDP) [33, 34, 40, 41] to calculate $R_d$ or ($1-R_d$) in terms of $d$-MPs or ($d$-1)-MCs, respectively. Moreover, both IET and SDP are also NP-hard.

However, the major drawback is that all existing algorithms in calculating $R_d$ are only available for fixed $d$, and decision-makers cannot choose the best $d$ under particular scenarios. It is insufficient to confirm the performance of MFNs and the roles of components for decision makers in making a correct decision. For instance, it is always for a decision maker to determine which values of $d$ can balance the most reliable, highest economic profit, and lowest cost to build transportation systems [7, 8], telecom systems [15], wireless sensor networks [4, 5, 6], smart grids [7], and the Internet of Things [1, 2].

Hence, it is necessary to propose an algorithm for determining the all-level reliability of MFN for all values of $d$ simultaneously, that is, calculate $R_1$, $R_2$, …, $R_{d\mathrm{MAX}}$ simultaneously. Owing to its importance and practicality, a novel direct problem is proposed to realize this.

The binary-addition-tree algorithm (BAT), first proposed by Yeh in [42], can determine all possible solutions, vectors, or combinations. Because BAT is based on binary addition, it is more efficient compared to the breadth-first search (BFS) and depth-first search (DFS), which are the basis of major search algorithms such as the branch-and-bound algorithm, decision tree [43], and others [44, 45]. Further, because BAT is simple, efficient, and memory saving, it has also been applied in different applications, that is, network reliability problems [42, 44, 45], rework problems [12], the propagation of computer viruses [46], and the spread of wildfire [47], and in addition, BAT has recently emerged as a new search method for finding all possible solutions or vectors.

Owing to the advantage of BAT [42], the goal of this study is to propose a BAT-based all-level direct algorithm to solve the proposed all-level MFN reliability problem, that is, calculate the all-level reliability $R_1$, $R_2$, …, $R_{d\mathrm{MAX}}$ simultaneously.

The remainder of this paper is organized as follows. All acronyms, notations, nomenclatures, and assumptions are introduced in Section II. A review of the BAT and the maximum-flow algorithm is presented in Section III. The proposed BAT-based algorithm is introduced formally in Section IV, including the difference between the indirect and direct algorithms and between the traditional BAT and the proposed BAT-based algorithm. The proposed algorithm determined the first connected vector, the pseudocode of the proposed algorithm, and an example to demonstrate the proposed algorithm. The performance of the proposed algorithm in solving the proposed problem was demonstrated on 12 benchmark networks in Section V. Finally, Section VI concludes the study.

## II. Acronyms, Notations, Nomenclatures, And Assumptions

All required acronyms, notations, assumptions, and nomenclatures are defined and provided here.

### *A. Acronyms*

BAT : binary-addition-tree algorithm [25]

MFN : multistate flow network

MP : minimal path

MC : minimal cut

*d*-MP : MP vector for demand *d*

*d*-MC : MC vector for demand *d*

DFS : depth-search-first algorithm

BFS : breadth-search-first algorithm

BBD : binary-decision-diagram

IET : inclusion-exclusion technology

UGFM : universal generating function methodology

## B. Notations

|•| : Element number of •

$n$ : Vertex number

$m$ : Edge number

$a_i$ : $i^{th}$ undirected edge

$V$ : Vertex set $V = \{1, 2, \ldots, n\}$ and $|V| = n$

$E$ : Edge set $E = \{a_1, a_2, \ldots, a_m\}$ and $|E| = m$

$e_{i,j}$ : $e_{i,j} = a_k \in E$ for one $k$ and $i, j \in V$

Pr(•) : Occurrence probability of event •

**D** : State distributions listing all states of a and its Pr($a$) for each edge $a$, for example, Table 1 is the state distribution of Fig. 1.

$G(V, E)$ : A graph constituted by $V$ and $E$. For example, Fig. 1 is the bridge network constituted by $E = \{a_1 = e_{1,2}, a_2 = e_{1,3}, a_3 = e_{2,3}, a_4 = e_{2,4}, a_5 = e_{3,4}\}$ and $V = \{1, 2, 3, 4\}$.

G($V$, $E$, **D**) : An MFN constituted by $G(V, E)$ and the state distributions **D**. For example, $G(V, E)$ in depicted Fig. 1 and its **D** is provided in Table 1 is an MFN $G(V, E, \mathbf{D})$.

$X$ : Multistate vector and its $k^{th}$ coordinate is the state of edge $a_k \in E$

$X(a_k)$ : State of the $k^{th}$ coordinate $a_k \in E$ in $X$ for $k = 1, 2, \ldots, m$

$\mathbf{D}_{max}(a_k)$ : The largest state of $a_k \in E$ for $k = 1, 2, \ldots, m$

Pr($X(a)$) : Occurrence probability of $a$ in vector $X$

Pr($X$) : $\prod_{k=1}^{m} \Pr(X(a_k))$

$R_d$ : Reliability for level d of an MFN

$r_d$ Probability that the exact *d* units of flow can be sent from nodes 1 to *n*

$G(X)$ : $G(X) = G(V, \{a\in E \mid \text{for all edge } a \text{ with } X(a) > 0\}, \mathbf{D})$

$F(X)$ : The maximum flow of $G(X)$

$d_{\text{MAX}}$ : $d_{\text{MAX}} = F(G)$ which is the maximum flow of $G(V, E, \mathbf{D})$

$X << Y$ : If vector *X* is obtained after *Y* in the BAT.

$X < Y$ : $X(a_i) < Y(a_i)$ for $i = 1, 2, \ldots, m$, e.g., (1, 0, 0, 0, 0) < (3, 1, 1, 1, 1).

$X \le Y$ : $X < Y$ and $X(a_i) = Y(a_i)$ for at least one *i*, e.g., (0, 1, 0, 1, 1) ≤ (3, 1, 1, 1, 1).

## C. Nomenclatures

1) MP/MC: An MP/MC is an edge subset such that the functioning/failure of all these edges ensures the functioning/failure of the network, and each of its proper subsets is not an MP/MC [20, 21]. For example, $\{a_1, a_3, a_5\}$ is an MC and an MP in Fig. 1.
2) *d*: random variable $d = 0, 1, 2, \ldots., d_{\text{MAX}}$ is the required amount of flow for a particular problem, and its distribution can be determined via constant prediction and observation.
3) *d*-MP: A vector *X* is a *d*-MP if and only if $F(X) = d$ and $F(X - o_j) < d$ for each $\mathbf{D}_{\max}(a_j)>0$ [48].
4) *d*-MP combination: A vector generated by the addition of one of (*d*−1)-MPs and one of 1-MPs [49].
5) $R_d$: The probability that at least *d* units of flows can be sent from vertices 1 and *n* for all values of *d*.
6) One-level reliability: The probability that at least d units of flows can be sent from vertices 1 and *n* for only a specific value of *d*.
7) All-levels reliability: $R_1$, $R_2$, …, and $R_{\text{dMAX}}$.
8) Reliability for all *d*: it is similar to all-level reliability except that it cannot calculate $R_d$ directly for all *d* at the same time
9) *d*-connected vector: A vector *X* is *d*-connected if $F(X) = d$ [45, 49].

### *D. Assumptions*

1) Each MFN is free of parallel edges or loops.
2) Each edge is undirected, and its state (capacity) is a zero or positive random variable followed by a predefined distribution.
3) States of different edges are statistically independent.
4) Each vertex is reliable perfectly.
5) The total flows into and from a node (not nodes 1 and *n*) are all equal, that is, the conservation law is followed.

## III. REVIEW OF BAT AND MAXIMUM FLOW

The proposed all-level direct multistate BAT was modified with the traditional (directed-arc) multistate BAT to search for each connected vector, say *X*, and used the maximum-flow algorithm to calculate *F*(*X*). Hence, both the traditional BAT and the maximum-flow algorithm are discussed in this section.

### *A. BAT*

The first BAT was proposed by Yeh [42] based on binary addition to solve binary-state directed network reliability problems. From experimental results, BAT algorithms outperform the BFS algorithm, for example, the universal generating function methodology (UGFM) [18], which is the main algorithm for solving the information network reliability problems [44], recursive SDP [33, 34]. Further, BAT outperforms depth-search-first algorithms (DFS), for example, the quick IET [39], which is the best-known algorithm for calculating MFN reliability after obtaining all *d*-MPs or (*d*-1)-MCs [42]. In addition, BAT is also better that binary-decision-diagram (BBD) [43, 45], which outperforms other algorithms in binary-state network reliability problems [45] when considering the running time and computer memory.

The original BAT can determine all possible solutions and starts from vector zero, say *X*, and adds one to *X* to update it iteratively. The entire procedure is backward from the last coordinate, that is, the $m^{\text{th}}$ coordinate, of *X*. If the current coordinate is zero, it needs to be updated to one, and we obtain a new solution *X*. However, if the current coordinate, say *i*, is one, it changes to zero and then moves to the next coordinate, that is, ($i$–1), to repeat the above procedure till the current coordinate is zero [42].

The original (backward) BAT pseudocode to find all solutions or vectors is provided in the following algorithm [25]:

**Algorithm:** BAT [42]

**Input:** A $m$-tuple zero vector.

**Output:** All solutions/vectors.

**STEP B0.** Let SUM = 0, $X$ = **0**, and $k = m$.

**STEP B1.** If $X(a_k) = 1$, let $X(a_k) = 0$, SUM = SUM − 1, and go to STEP B4.

**STEP B2.** Let $X(a_k) = 1$ and SUM = SUM + 1.

**STEP B3.** Let $k = k - 1$ and go to STEP B1 if $k > 1$.

**STEP B4.** If SUM = $m$, halt; otherwise, let $k = m$ and go to STEP B1.

STEP B0 initializes the values of SUM, which is used to determine when to stop, the current vector $X$ to be vector zero, and the current coordinate $k$ to be $m$ (backward of the backward BAT). In STEP B1, based on the binary addition, the current coordinate is changed to 0 if it is 1 and then go to STEP B4, for example, (1, 0, 1) is updated to (1, 1, 0). STEP B2 deals with other cases, that is, if the current coordinate is 0, it changes it to 1, and a new binary vector is generated, for example, (1, 1, 0) is updated to (1, 1, 1). Finally, STEP B4 represents the stopping criterion based on the SUM value.

The number of all found $X$ is $2^m$ if BAT starts with an $m$-tuple vector zero in STEP B0 with a time complexity of $O(2^{m+1})$, as proven in [50]. Moreover, the main total memory required for BAT is $O(m)$ only because the current vector $X$ is used repeatedly. However, after obtaining a new vector, we need to calculate its related values, for example, the reliability in this study, probability, profit, cost, time, or any predefined function [42].

From the above pseudocode, it is evident that BAT is simple to understand, easy to code, efficient to execute, economical in memory space, and convenient to make-to-fit [42, 44, 45, 46, 47, 50]. The following table shows the manner in which the BAT is implemented to calculate the exact reliability of the binary-state network shown in Fig 1, where $|E| = 5$ and $p_i = 1 - q_i$ is the reliability of arc $a_i$ in **D** for all $a_i \in E$.

**TABLE 2.** COMPLETE RESULTS OBTAINED FROM THE BAT ON FIG. 1.

| Iterative | $X$ | SUM | Connected? | $R(X)$ |
|---|---|---|---|---|
| 1 | (0, 0, 0, 1, 1) | 2 | N | |
| 2 | (0, 0, 1, 0, 0) | 1 | N | |
| 3 | (0, 0, 1, 0, 1) | 2 | N | |
| 4 | (0, 0, 1, 1, 0) | 2 | N | |
| 5 | (0, 0, 1, 1, 1) | 3 | N | |
| 6 | (0, 1, 0, 0, 0) | 1 | N | |
| 7 | (0, 1, 0, 0, 1) | 2 | Y | $q_1p_2q_3q_4p_5$ |
| 8 | (0, 1, 0, 1, 0) | 2 | N | |
| 9 | (0, 1, 0, 1, 1) | 3 | Y | $q_1p_2q_3p_4p_5$ |
| 10 | (0, 1, 1, 0, 0) | 2 | N | |
| 11 | (0, 1, 1, 0, 1) | 3 | Y | $q_1p_2p_3q_4p_5$ |
| 12 | (0, 1, 1, 1, 0) | 3 | N | |
| 13 | (0, 1, 1, 1, 1) | 4 | Y | $q_1p_2p_3p_4p_5$ |
| 14 | (1, 0, 0, 0, 0) | 1 | N | |
| 15 | (1, 0, 0, 0, 1) | 2 | N | |
| 16 | (1, 0, 0, 1, 0) | 2 | Y | $p_1q_2q_3p_4q_5$ |
| 17 | (1, 0, 0, 1, 1) | 3 | Y | $p_1q_2q_3p_4p_5$ |
| 18 | (1, 0, 1, 0, 0) | 2 | N | |
| 19 | (1, 0, 1, 0, 1) | 3 | Y | $p_1q_2p_3q_4p_5$ |
| 20 | (1, 0, 1, 1, 0) | 3 | Y | $p_1q_2p_3p_4q_5$ |
| 21 | (1, 0, 1, 1, 1) | 4 | Y | $p_1q_2p_3p_4p_5$ |
| 22 | (1, 1, 0, 0, 0) | 2 | N | |
| 23 | (1, 1, 0, 0, 1) | 3 | Y | $p_1p_2q_3q_4p_5$ |
| 24 | (1, 1, 0, 1, 0) | 3 | Y | $p_1p_2q_3p_4q_5$ |
| 25 | (1, 1, 0, 1, 1) | 4 | Y | $p_1p_2q_3p_4p_5$ |
| 26 | (1, 1, 1, 0, 0) | 3 | N | |
| 27 | (1, 1, 1, 0, 1) | 4 | Y | $p_1p_2p_3q_4p_5$ |
| 28 | (1, 1, 1, 1, 0) | 4 | Y | $p_1p_2p_3p_4q_5$ |
| 29 | (1, 1, 1, 1, 1) | 5 | Y | $p_1p_2p_3p_4p_5$ |

### *B. Maximum-Flow Algorithm*

The maximum-flow algorithm can calculate the maximum flow in a related subnetwork corresponding to a vector. Numerous maximum-flow algorithms exist in the literature [51]. In this study, we adapted the simplest one by repeating a simple path, say $P$, using DFS in the residual subnetwork. Thereafter, we find the minimum state, say $f_1$, among the states of all edges in $P$, and subtract the states of all edges in $P$ by $f_1$ to obtain a new residual subnetwork. Consequently, the above process is repeated and all $f_i$ for all $i$ are summed to obtain the maximum flow.

The pseudocode of the algorithm to calculate the maximum flow in the proposed all-level direct multistate BAT is delivered as follows:

**Algorithm:** The Maximum-flow Algorithm

**Input:** A vector $X$

**Output:** The maximum flow $F(X)$.

**STEP M0.** Let $F^* = 0$.

**STEP M1.** Find a simple path, say $P \in E$, from nodes 1 to $n$. If there is no such path, go to STEP M4.

**STEP M2.** Let $f = \text{Min}\{\ F(a) \mid \text{for all} \in\ P\ \}$.

**STEP M3.** Let $F^* = F^* + f$, $F(a) = F(a) - f$ for all $a \in\ P$ and go to STEP M0.

**STEP M4.** $F(X) = F^*$ and halt.

For example, let $X = (1, 1, 1, 2, 0)$ in Fig. 1, and the corresponding subnetwork $G(X)$ is depicted in Fig. 2.

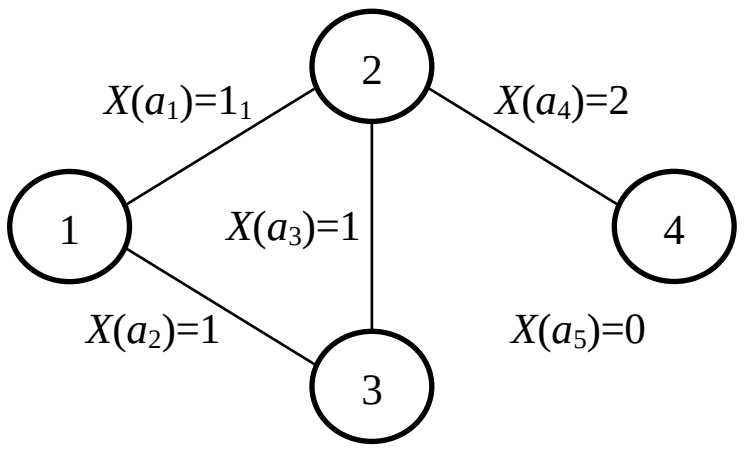

**Figure 2.** The $G(X)$ and $X = (1, 1, 1, 2, 0)$.

There is a simple path from nodes 1 to 4, that is, $P_1 = \{a_1, a_4\}$, as shown by the dashed line in Fig. 3(1), and the minimum state in such a path is $\text{Min}\{X(a_1) = 1, X(a_4) = 2\} = 1$. Hence, the state of all arcs in $P_1$ must be subtracted by 1, as shown in Fig. 3(2).

Similarly, there is another simple path from nodes 1 to 4 in the residual network, that is, $P_2 = \{a_2, a_3, a_4\}$. The minimum state of all arcs in $P_2$ is $\text{Min}\{X(a_2) = 1, X(a_3) = 1, X(a_4) = 1\} = 1$, as shown in Fig. 3(3). In addition, there is no other simpler path in the residual network after subtracting the state of arcs in $P_2$ by 1. Hence, the maximum flow was $1 + 1 = 2$.

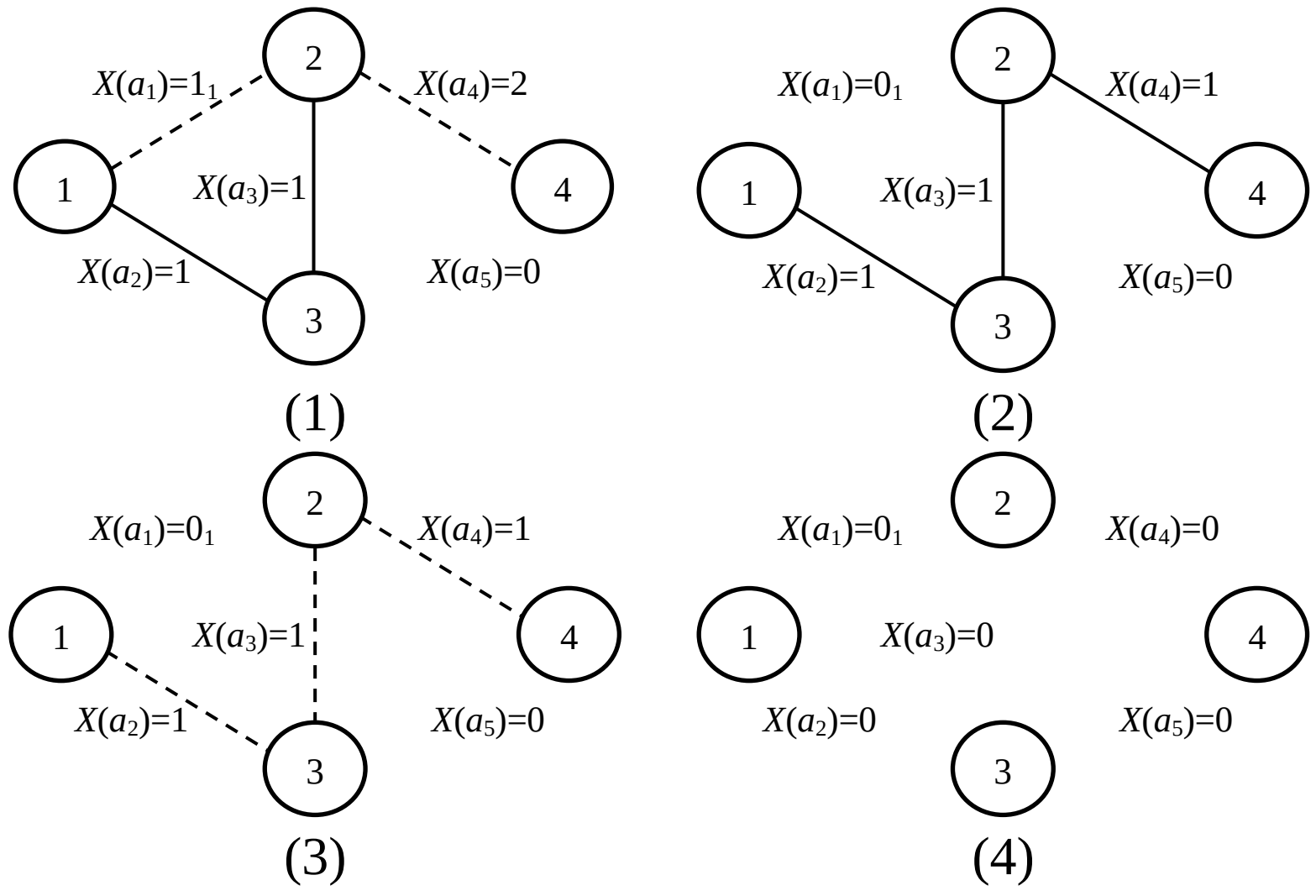

**Figure 3.** The related subnetwork in calculating $F$(X).

The complete process of the above process is explained in Table 3.

**TABLE 3.** CALCULATE $F(X)$ FOR $X = (1, 1, 1, 2, 0)$ FOR FIG. 1.

| $i$ | $P_i$ | Min{$X(a)$ \| for all $a \in P_i$ } |
|---|---|---|
| 1 | $\{a_1, a_4\}$ | 1 |
| 2 | $\{a_2, a_3, a_4\}$ | 1 |
| SUM | | 2 |

## IV. PROPOSED ALL-LEVELS MULTI-BAT

The proposed all-level direct multistate BAT for computing the all-level reliability of MFNs is presented, including a discussion of the differences between the proposed problem and the traditional $d$-MP for all $d$ problems and between the proposed BAT and the traditional BAT in Section 4.1. The first connected vector to reduce the runtime on finding the disconnected vector is presented in Section 4.2, the pseudocode and time complexity of the all-level BAT in Section 4.3, and finally, an example to demonstrate the proposed all-levels BAT in Section 4.4.

### *A. Major Differences between the Proposed Direct BAT*

In the traditional $d$-MP-for-all-$d$ algorithm [49], $d$-MPs can only be found for all possible $d$ values and reliability for all levels cannot be calculated directly. It states that all ($d$+1)-MPs are elements of the set that constitutes one $d$-MP plus one 1-MP. This is trivial, and its proof is omitted here [49].

**Algorithm:** the $d$-MP-for-all-$d$ Algorithm [49]

**Input:** MFN $G(V, E, \mathbf{D})$ with source node 1 and sink node $n$.

**Output:** The reliability $R_d$ of $G(V, E, \mathbf{D})$ for $d = 1, 2, \ldots, F(G)$.

**STEP D0.** Find $P_1$ = {all MPs}, and let $d = 1$.

**STEP D1.** Find $P_{(d+1)} = \{ p \mid p = p_1 + p_d$ for all $p_d \in P_d$, $p_1 \in P_1$, and $p \notin P_{(d+1)}$ if $p^* \le p$ for all $p^* \in P_{(d+1)}\}$.

**STEP D2.** Calculate $R_{(d+1)}$ in terms of all ($d$+1)-MPs [18-26] using IET or SDP.

**STEP D3.** If $d < F(G)$, let $d = d + 1$ and go to STEP D1; otherwise, it is halted.

A ($d$+1)-MP combination is a vector generated by the addition of one of the $d$-MPs and one of the 1-MPs; for example, $p$ in STEP D1. A combination $X$ is a redundant combination if $F(X) = F(X^*) = d$ and $X^* \le X$ [49]. Further, if $X = X^*$ and $F(X) = F(X^*) = d$, $X$ and $X^*$ are generated from different $d$-MPs and 1-MPs, and $X$ and $X^*$ are duplicates.

A ($d$+1)-MP combination is a real ($d$+1)-MP if it is not a redundant or duplicate combination [49]. Hence, in STEP D1, each redundant and duplicate combination, both of which are NP-hard, must be removed. From STEP D2, after finding all ($d$+1)-MPs, the next step to calculate the reliability is to implement the IET or the SDP, both of which are also NP-hard. However, the proposed all-level direct multistate BAT can have all levels of reliability directly without using IET or SDP and also without the need to remove redundant and duplicate combinations.

It should be noted that to date there is no algorithm that can calculate the all-level MFN reliability directly simultaneously.

In addition, there are two major differences between the proposed all-level multistate BAT and the traditional BAT.

1. The proposed BAT determines the first connected vector, $X_{FC}$, in a novel way such that the proposed algorithm starts to update from $X_{FC}$, in contrast to the other BATs, that start from vector zero.
2. Immediately after a connected vector, $X$, is formed, the proposed BAT needs to calculate the

maximal flow $F(X)$.

Moreover, in the proposed BAT, there is no need to change each undirected edge to two directed edges in opposite directions.

*B. First Connected Vector*

The idea of the first connected vector $X_{\mathrm{FC}}$ was first proposed in [45] to reduce the number of disconnected vectors in binary-state networks. The vector $X_{\mathrm{FC}}$ is a special vector, and it is also the first connected vector such that $X$ is a disconnected vector if $X << X_{\mathrm{FC}}$ in the BAT update procedure. For example, $X_{\mathrm{FC}} = (1, 0, 0, 1, 0)$ in Fig. 1, because $X$ is disconnected for all $X << X_{\mathrm{FC}}$.

If the values of any coordinate $i$ with $X_{\mathrm{FC}}(a_i) > 0$ are reduced to 0, the new $X_{\mathrm{FC}}$, $X_{\mathrm{new,FC}}$, is disconnected because $X_{\mathrm{new,FC}} << X_{\mathrm{FC}}$, that is, (0, 0, 0, 0, 0), (1, 0, 0, 0, 0), and (0, 0, 0, 1, 0) are all disconnected, as shown in Fig. 1. Conversely, if $X$ is the disconnected vector immediately before $X_{FC}$, $X$ is connected after adding one to $X$ using binary addition. The above concept is the core of finding the $X_{\mathrm{FC}}$.

Compared to the $X_{\mathrm{FC}}$ in [45], the method to obtain $X_{\mathrm{FC}}$ is revised here, and only the definition of $X_{\mathrm{FC}}$ proposed in [45] is retained. The $X_{\mathrm{FC}}$ is extended to the MFN, and a more efficient algorithm is proposed to find the $X_{\mathrm{FC}}$ simply based on the well-known shortest path. If $X = X_{\mathrm{FC}}$ in a binary-state network $G(V, E, D)$, $X = X_{\mathrm{FC}}$ in an MFN by changing each binary state to a multistate in MFN because $G(X_{\mathrm{FC}})$ is an MP. Conversely, $X_{\mathrm{FC}} \neq X$ if $G(X)$ includes more than one MP. The reason is also based on the definition of $X_{\mathrm{FC}}$, that is, if $X$ is the disconnected vector immediately before $X_{FC}$, $X$ is connected after adding one to $X$ using the binary addition. The above concept is the core in finding $X_{\mathrm{FC}}$, and the $X_{\mathrm{FC}}$ by using the following code:

**Algorithm FIND_$X_{\mathrm{FC}}$**

**Input:** MFN $G(V, E, \mathbf{D})$ with source node 1 and sink node $n$.

**Output:** The $X_{\mathrm{FC}}$.

**STEP F0.** Let the distance of $a_i$ be $2^{(i-1)}$ for $i = 1, 2, \ldots, m$.

**STEP F1.** Find the shortest path, $P$, in update network.

**STEP F2.** Let $X_{FC}(a_i) = 1$ if $a_i \in P$ for $i = 1, 2, \ldots, m$.

Any vector in a binary-state network can be transferred into a unique number, as discussed in [42]. For example, the number corresponding to (1, 0, 0, 1, 0) is $1\times2^0 + 0\times2^1 + 0\times2^2 + 0\times2^3 + 1\times2^4 = 17$. Hence, STEP F0 transfers each edge label to a distance power of 2. STEP F1 simply finds the shortest path $P$ based on these distances transferred from the edge labels in STEP F0. Further, STEP F2 transferred each distance in the shortest path back to the corresponding label to obtain an $X_{FC}$. In addition, the shortest path $P$ is also an MP and the first MP in the BAT procedure.

For example, Fig. 4. is the corresponding network after implementing the STEP F0 in Fig. 1. The shortest path is $\{e_{1,2}, e_{2,4}\}$ with a distance of 9. Hence, $X_{FC}$ = (1, 0, 0, 1, 0).

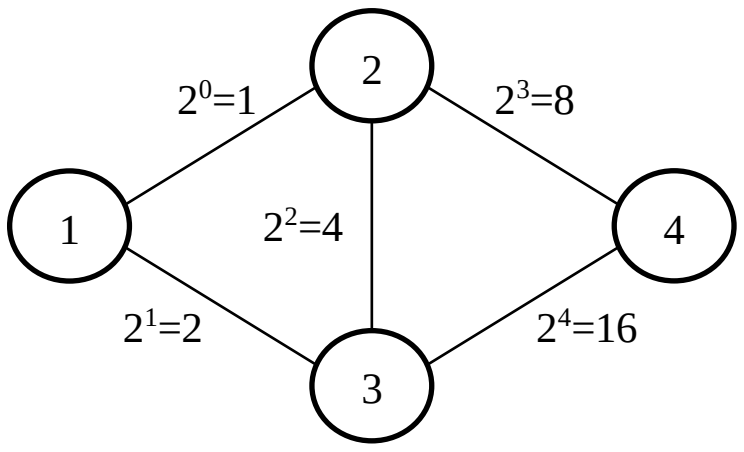

**Figure 4.** The new network after transferring each label into a distance in Fig. 1.

After finding the $X_{FC}$, there are 28 disconnected vectors that can be skipped as shown below:

**TABLE 4.** ALL 28 DISCONNECTED VECTORS ARE SKIPPED WITHOUT NEEDING TO FIND BASED ON $X_{FC}$.

| Order in BAT | $x_1$ | $x_2$ | $x_3$ | $x_4$ | $x_5$ |
|---|---|---|---|---|---|
| 1 | 0 | 0 | 0 | 0 | 0 |
| 2 | 1 | 0 | 0 | 0 | 0 |
| 3 | 2 | 0 | 0 | 0 | 0 |
| 4 | 0 | 1 | 0 | 0 | 0 |
| 5 | 1 | 1 | 0 | 0 | 0 |
| 6 | 2 | 1 | 0 | 0 | 0 |
| 7 | 0 | 2 | 0 | 0 | 0 |
| 8 | 1 | 2 | 0 | 0 | 0 |
| 9 | 2 | 2 | 0 | 0 | 0 |
| 10 | 0 | 0 | 1 | 0 | 0 |
| 11 | 1 | 0 | 1 | 0 | 0 |
| 12 | 2 | 0 | 1 | 0 | 0 |
| 13 | 0 | 1 | 1 | 0 | 0 |
| 14 | 1 | 1 | 1 | 0 | 0 |
| 15 | 2 | 1 | 1 | 0 | 0 |
| 16 | 0 | 2 | 1 | 0 | 0 |
| 17 | 1 | 2 | 1 | 0 | 0 |
| 18 | 2 | 2 | 1 | 0 | 0 |
| 19 | 0 | 0 | 2 | 0 | 0 |
| 20 | 1 | 0 | 2 | 0 | 0 |
| 21 | 2 | 0 | 2 | 0 | 0 |
| 22 | 0 | 1 | 2 | 0 | 0 |
| 23 | 1 | 1 | 2 | 0 | 0 |
| 24 | 2 | 1 | 2 | 0 | 0 |
| 25 | 0 | 2 | 2 | 0 | 0 |
| 26 | 1 | 2 | 2 | 0 | 0 |
| 27 | 2 | 2 | 2 | 0 | 0 |
| 28 | 0 | 0 | 0 | 1 | 0 |

### *C. Pseudocode and Time Complexity*

The pseudocode of the proposed all-level direct multistate BAT is presented below.

**Algorithm:** All-Levels Direct Multistate BAT

**Input:** An MFN $G(V, E, \mathbf{D})$

**Output:** Calculate the reliability $R_d$ for $d = 1, 2, \ldots, F(G)$ simultaneously.

**STEP 0.** Find $X_{\text{FC}}$ based on Section IVB, and let $X = X_{\text{FC}}$, $i = 1$, and $R_d = 0$ for $d = 1, 2, \ldots, F(G)$.

**STEP 1.** If $X(a_i) = U(a_i)$, then $X(a_i) = 0$. Otherwise, go to STEP 3.

**STEP 2.** If $i < m$, let $i = i + 1$ and go to STEP 1. Otherwise, halt.

**STEP 3.** Let $X(a_i) = X(a_i) + 1$.

**STEP 4.** Calculate $F(X)$, say $f$, and let $R_f = R_f + \Pr(X)$ if $F(X) > 0$, $i = 1$, and go to STEP 1.

There are only five lines in the above pseudocode, and $X$ is used repeatedly. Hence, the proposed all-level direct multistate BAT is very easy to code, understand, saves computer memory, and implement.

STEP 0 initializes all values, for example, $X$, $i$, and $R_d$ for all $d = 1, 2, \ldots, F(G) = d_{\text{MAX}}$. STEPs 1 – 3, are based on the multistate BAT discussed in Section 3. The major difference is to halt the program if the current $X$ is back to the $m$-tuple vector zero, that is, $i = m + 1$ in STEP 3. Finally, STEP 4 computes the maximum flow, $f$, calculates $\Pr(X)$, and adds $\Pr(X)$ to $R_f$.

The time complexity of the multistate BAT is discussed below because it is a key index for judging the efficiency of a new algorithm. Let each edge, $a_i$ have $s_i$ states, that is, 0, 1, …, $s_{i\text{-}1}$. There are $m$ coordinates in each vector; hence, the complete number of all vectors (without duplicates) generated from the BAT is $s_1 \times s_2 \times \ldots \times s_m$. To update the current vector $X$ in the forward multistate BAT, we need to determine the first coordinate, $i$, with value 0 from the 1st coordinate to the $m$th coordinate, that is, $O(m)$ in the worst case for each update. Thus, the total time complexity is $O(ms_1s_2)$ for multistate BAT.

Furthermore, in the proposed all-level BAT, we need to consider STEP 4, which is the major difference between the proposed BAT and the multistate BAT. The maximum flow can be calculated in $O(mn)$ [51], and $O(m)$ is required to calculate Pr($X$) in STEP 4, that is, $O(mn) + O(m) \approx O(mn)$ [50]. Consequently, the time complexity of the proposed all-level BAT is $O(nm^2s_1s_2\ldots s_m)$.

*D. Example*

Regardless of whether the network is binary-state, multistate, one-level, for all $d$, or the proposed all-levels, all are NP-hard and # P-hard, that is, it is complex and tedious to calculate all types of MFN reliability problems. Hence, to better demonstrate the proposed all-level direct multistate BAT, the all-level reliability of a smaller-size network is implemented. Thus, the bridge network shown in Fig. 1 is adopted to demonstrate that readers can understand the proposed all-level reliability of the MFN problem and the proposed all-level direct multistate BAT quickly.

The proposed all-level direct multistate BAT is implemented in the bridge network, which has five edges in a backward manner, starting from the first coordinate to the initial 5-tuple zero vector, that is, $X$ = (0, 0, 0, 0, 0). The step-by-step procedure of the proposed algorithm is as follows.

**STEP 0.** Find $X_{FC}$ = (1, 0, 0, 1, 0) based on Section IVB and let $X = X_{FC}$ = (1, 0, 0, 1, 0), $i$ = 1, and $R_1$ = Pr($X$) = $0.375\times10^6$, $R_d$ = 0 for $d$ = 2, 3, $F(G)$ = 4.

**STEP 1.** Because $X(a_1) = 1 < U(a_1) = 2$, go to STEP 3.

**STEP 3.** Let $X(a_1) = X(a_1) + 1 = 2$, that is, $X$ = (2, 0, 0, 1, 0).

**STEP 4.** Because $F(X) = 1$, let $R_1 = R_1 + \Pr(X) = 1.688 \times 10^6$, $i = 1$, and go to STEP 1.

:

:

**STEP 3.** Let $X(a_1) = X(a_1) + 1 = 1$, that is, $X = (1, 2, 2, 2, 2)$.

**STEP 4.** Because $F(X) = 3$, let $R_3 = R_3 + \Pr(X) = 0.384884446$, $i = 1$, and go to STEP 1.

**STEP 1.** Because $X(a_1) = 1 < U(a_1) = 2$, go to STEP 3.

**STEP 3.** Let $X(a_1) = X(a_1) + 1 = 2$, that is, $X = (2, 2, 2, 2, 2)$.

**STEP 4.** Because $F(X) = 4$, let $R_4 = R_4 + \Pr(X) =0.429671250$, $i = 1$, and go to STEP 1.

**STEP 1.** Because $X(a_1) = U(a_1) = 2$, let $X(a_1) = 0$ and go to STEP 2.

**STEP 2.** Because $i = 1 < m = 5$, let $i = i + 1 = 2$ and go to STEP 1.

The results obtained using the proposed BAT are provided in Tables 5 and 6. Each coordinate can be 0, 1, 2, and 3, that is, there are $4^5 = 243$ different vectors if $X_{FC}$ is not found. However, after determining the $X_{FC}$, the number of all vectors was reduced to 225.

**TABLE 5.** ALL 225 VECTORS OBTAINED FROM THE PROPOSED ALL-LEVELS BAT.

| $i$ | $X_i$ | $\Pr(X_i)\times10^6$ | $d$ | $i$ | $X_i$ | $\Pr(X_i)\times10^6$ | $d$ | $i$ | $X_i$ | $\Pr(X_i)\times10^6$ | $d$ | $i$ | $X_i$ | $\Pr(X_i)\times10^6$ | $d$ |
|---|---|---|---|---|---|---|---|---|---|---|---|---|---|---|---|
| 1 | (1,0,0,1,0) | 0.375 | 1 | 56 | (2,0,0,0,1) | 0 | | 111 | (0,1,0,2,1) | 5.625 | 1 | 166 | (1,1,0,1,2) | 27.000 | 3 |
| 2 | (2,0,0,1,0) | 1.313 | 1 | 57 | (0,1,0,0,1) | 0.750 | 1 | 112 | (1,1,0,2,1) | 11.250 | 3 | 167 | (2,1,0,1,2) | 94.500 | 3 |
| 3 | (0,1,0,1,0) | 0 | | 58 | (1,1,0,0,1) | 1.500 | 1 | 113 | (2,1,0,2,1) | 39.375 | 3 | 168 | (0,2,0,1,2) | 114.750 | 2 |
| 4 | (1,1,0,1,0) | 0.750 | 1 | 59 | (2,1,0,0,1) | 5.250 | 1 | 114 | (0,2,0,2,1) | 47.813 | 1 | 169 | (1,2,0,1,2) | 229.500 | 3 |
| 5 | (2,1,0,1,0) | 2.625 | 1 | 60 | (0,2,0,0,1) | 6.375 | 1 | 115 | (1,2,0,2,1) | 95.625 | 3 | 170 | (2,2,0,1,2) | 803.250 | 3 |
| 6 | (0,2,0,1,0) | 0 | | 61 | (1,2,0,0,1) | 12.750 | 1 | 116 | (2,2,0,2,1) | 334.688 | 3 | 171 | (0,0,1,1,2) | 0 | |
| 7 | (1,2,0,1,0) | 6.375 | 1 | 62 | (2,2,0,0,1) | 44.625 | 1 | 117 | (0,0,1,2,1) | 0 | | 172 | (1,0,1,1,2) | 256.500 | 2 |
| 8 | (2,2,0,1,0) | 22.313 | 1 | 63 | (0,0,1,0,1) | 0 | | 118 | (1,0,1,2,1) | 106.875 | 3 | 173 | (2,0,1,1,2) | 897.750 | 2 |
| 9 | (0,0,1,1,0) | 0 | | 64 | (1,0,1,0,1) | 14.250 | 1 | 119 | (2,0,1,2,1) | 374.063 | 3 | 174 | (0,1,1,1,2) | 256.500 | 3 |
| 10 | (1,0,1,1,0) | 7.125 | 1 | 65 | (2,0,1,0,1) | 49.875 | 1 | 120 | (0,1,1,2,1) | 106.875 | 2 | 175 | (1,1,1,1,2) | 513.000 | 2 |
| 11 | (2,0,1,1,0) | 24.938 | 1 | 66 | (0,1,1,0,1) | 14.250 | 1 | 121 | (1,1,1,2,1) | 213.750 | 3 | 176 | (2,1,1,1,2) | 1795.500 | 2 |
| 12 | (0,1,1,1,0) | 7.125 | 1 | 67 | (1,1,1,0,1) | 28.500 | 1 | 122 | (2,1,1,2,1) | 748.125 | 3 | 177 | (0,2,1,1,2) | 2180.250 | 3 |
| 13 | (1,1,1,1,0) | 14.250 | 1 | 68 | (2,1,1,0,1) | 99.750 | 1 | 123 | (0,2,1,2,1) | 908.438 | 2 | 178 | (1,2,1,1,2) | 4360.500 | 2 |
| 14 | (2,1,1,1,0) | 49.875 | 1 | 69 | (0,2,1,0,1) | 121.125 | 1 | 124 | (1,2,1,2,1) | 1816.875 | 3 | 179 | (2,2,1,1,2) | 15261.750 | 2 |
| 15 | (0,2,1,1,0) | 60.563 | 1 | 70 | (1,2,1,0,1) | 242.250 | 1 | 125 | (2,2,1,2,1) | 6359.063 | 3 | 180 | (0,0,2,1,2) | 0 | |
| 16 | (1,2,1,1,0) | 121.125 | 1 | 71 | (2,2,1,0,1) | 847.875 | 1 | 126 | (0,0,2,2,1) | 0 | | 181 | (1,0,2,1,2) | 1080.000 | 2 |
| 17 | (2,2,1,1,0) | 423.938 | 1 | 72 | (0,0,2,0,1) | 0 | | 127 | (1,0,2,2,1) | 450.000 | 3 | 182 | (2,0,2,1,2) | 3780.000 | 3 |
| 18 | (0,0,2,1,0) | 0 | | 73 | (1,0,2,0,1) | 60.000 | 1 | 128 | (2,0,2,2,1) | 1575.000 | 3 | 183 | (0,1,2,1,2) | 1080.000 | 3 |
| 19 | (1,0,2,1,0) | 30.000 | 1 | 74 | (2,0,2,0,1) | 210.000 | 1 | 129 | (0,1,2,2,1) | 450.000 | 2 | 184 | (1,1,2,1,2) | 2160.000 | 2 |
| 20 | (2,0,2,1,0) | 105.000 | 1 | 75 | (0,1,2,0,1) | 60.000 | 1 | 130 | (1,1,2,2,1) | 900.000 | 3 | 185 | (2,1,2,1,2) | 7560.000 | 3 |
| 21 | (0,1,2,1,0) | 30.000 | 1 | 76 | (1,1,2,0,1) | 120.000 | 1 | 131 | (2,1,2,2,1) | 3150.000 | 3 | 186 | (0,2,2,1,2) | 9180.001 | 3 |
| 22 | (1,1,2,1,0) | 60.000 | 1 | 77 | (2,1,2,0,1) | 420.000 | 1 | 132 | (0,2,2,2,1) | 3825.000 | 3 | 187 | (1,2,2,1,2) | 18360.001 | 2 |
| 23 | (2,1,2,1,0) | 210.000 | 1 | 78 | (0,2,2,0,1) | 510.000 | 1 | 133 | (1,2,2,2,1) | 7650.001 | 3 | 188 | (2,2,2,1,2) | 64260.003 | 3 |
| 24 | (0,2,2,1,0) | 255.000 | 1 | 79 | (1,2,2,0,1) | 1020.000 | 1 | 134 | (2,2,2,2,1) | 26775.002 | 3 | 189 | (0,0,0,2,2) | 0 | |
| 25 | (1,2,2,1,0) | 510.000 | 1 | 80 | (2,2,2,0,1) | 3570.000 | 1 | 135 | (0,0,0,0,2) | 0 | | 190 | (1,0,0,2,2) | 67.500 | 2 |
| 26 | (2,2,2,1,0) | 1785.000 | 1 | 81 | (0,0,0,1,1) | 0 | | 136 | (1,0,0,0,2) | 0 | | 191 | (2,0,0,2,2) | 236.250 | 2 |
| 27 | (0,0,0,2,0) | 0 | | 82 | (1,0,0,1,1) | 1.125 | 1 | 137 | (2,0,0,0,2) | 0 | | 192 | (0,1,0,2,2) | 67.500 | 2 |
| 28 | (1,0,0,2,0) | 1.875 | 2 | 83 | (2,0,0,1,1) | 3.938 | 1 | 138 | (0,1,0,0,2) | 9.000 | 2 | 193 | (1,1,0,2,2) | 135.000 | 4 |
| 29 | (2,0,0,2,0) | 6.563 | 2 | 84 | (0,1,0,1,1) | 1.125 | 1 | 139 | (1,1,0,0,2) | 18.000 | 2 | 194 | (2,1,0,2,2) | 472.500 | 4 |
| 30 | (0,1,0,2,0) | 0 | | 85 | (1,1,0,1,1) | 2.250 | 2 | 140 | (2,1,0,0,2) | 63.000 | 2 | 195 | (0,2,0,2,2) | 573.750 | 2 |
| 31 | (1,1,0,2,0) | 3.750 | 2 | 86 | (2,1,0,1,1) | 7.875 | 2 | 141 | (0,2,0,0,2) | 76.500 | 2 | 196 | (1,2,0,2,2) | 1147.500 | 4 |
| 32 | (2,1,0,2,0) | 13.125 | 2 | 87 | (0,2,0,1,1) | 9.563 | 1 | 142 | (1,2,0,0,2) | 153.000 | 2 | 197 | (2,2,0,2,2) | 4016.250 | 4 |
| 33 | (0,2,0,2,0) | 0 | | 88 | (1,2,0,1,1) | 19.125 | 2 | 143 | (2,2,0,0,2) | 535.500 | 2 | 198 | (0,0,1,2,2) | 0 | |
| 34 | (1,2,0,2,0) | 31.875 | 2 | 89 | (2,2,0,1,1) | 66.938 | 2 | 144 | (0,0,1,0,2) | 0 | | 199 | (1,0,1,2,2) | 1282.500 | 3 |
| 35 | (2,2,0,2,0) | 111.563 | 2 | 90 | (0,0,1,1,1) | 0 | | 145 | (1,0,1,0,2) | 171.000 | 1 | 200 | (2,0,1,2,2) | 4488.750 | 3 |
| 36 | (0,0,1,2,0) | 0 | | 91 | (1,0,1,1,1) | 21.375 | 2 | 146 | (2,0,1,0,2) | 598.500 | 1 | 201 | (0,1,1,2,2) | 1282.500 | 3 |
| 37 | (1,0,1,2,0) | 35.625 | 2 | 92 | (2,0,1,1,1) | 74.813 | 2 | 147 | (0,1,1,0,2) | 171.000 | 2 | 202 | (1,1,1,2,2) | 2565.000 | 3 |
| 38 | (2,0,1,2,0) | 124.688 | 2 | 93 | (0,1,1,1,1) | 21.375 | 2 | 148 | (1,1,1,0,2) | 342.000 | 1 | 203 | (2,1,1,2,2) | 8977.500 | 3 |
| 39 | (0,1,1,2,0) | 35.625 | 1 | 94 | (1,1,1,1,1) | 42.750 | 2 | 149 | (2,1,1,0,2) | 1197.000 | 1 | 204 | (0,2,1,2,2) | 10901.250 | 3 |
| 40 | (1,1,1,2,0) | 71.250 | 2 | 95 | (2,1,1,1,1) | 149.625 | 2 | 150 | (0,2,1,0,2) | 1453.500 | 2 | 205 | (1,2,1,2,2) | 21802.500 | 3 |
| 41 | (2,1,1,2,0) | 249.375 | 2 | 96 | (0,2,1,1,1) | 181.688 | 2 | 151 | (1,2,1,0,2) | 2907.000 | 1 | 206 | (2,2,1,2,2) | 76308.748 | 3 |
| 42 | (0,2,1,2,0) | 302.813 | 1 | 97 | (1,2,1,1,1) | 363.375 | 2 | 152 | (2,2,1,0,2) | 10174.500 | 1 | 207 | (0,0,2,2,2) | 0 | |
| 43 | (1,2,1,2,0) | 605.625 | 2 | 98 | (2,2,1,1,1) | 1271.813 | 2 | 153 | (0,0,2,0,2) | 0 | | 208 | (1,0,2,2,2) | 5400.000 | 3 |
| 44 | (2,2,1,2,0) | 2119.688 | 2 | 99 | (0,0,2,1,1) | 0 | | 154 | (1,0,2,0,2) | 720.000 | 1 | 209 | (2,0,2,2,2) | 18900.000 | 4 |
| 45 | (0,0,2,2,0) | 0 | | 100 | (1,0,2,1,1) | 90.000 | 2 | 155 | (2,0,2,0,2) | 2520.000 | 2 | 210 | (0,1,2,2,2) | 5400.000 | 3 |
| 46 | (1,0,2,2,0) | 150.000 | 2 | 101 | (2,0,2,1,1) | 315.000 | 2 | 156 | (0,1,2,0,2) | 720.000 | 2 | 211 | (1,1,2,2,2) | 10800.000 | 3 |
| 47 | (2,0,2,2,0) | 525.000 | 2 | 102 | (0,1,2,1,1) | 90.000 | 2 | 157 | (1,1,2,0,2) | 1440.000 | 1 | 212 | (2,1,2,2,2) | 37800.000 | 4 |
| 48 | (0,1,2,2,0) | 150.000 | 1 | 103 | (1,1,2,1,1) | 180.000 | 2 | 158 | (2,1,2,0,2) | 5040.000 | 2 | 213 | (0,2,2,2,2) | 45900.001 | 4 |
| 49 | (1,1,2,2,0) | 300.000 | 2 | 104 | (2,1,2,1,1) | 630.000 | 2 | 159 | (0,2,2,0,2) | 6120.000 | 2 | 214 | (1,2,2,2,2) | 91800.003 | 3 |
| 50 | (2,1,2,2,0) | 1050.000 | 2 | 105 | (0,2,2,1,1) | 765.000 | 2 | 160 | (1,2,2,0,2) | 12240.001 | 1 | 215 | (2,2,2,2,2) | 321300.000 | 4 |
| 51 | (0,2,2,2,0) | 1275.000 | 2 | 106 | (1,2,2,1,1) | 1530.000 | 2 | 161 | (2,2,2,0,2) | 42840.001 | 2 | | | | |
| 52 | (1,2,2,2,0) | 2550.000 | 2 | 107 | (2,2,2,1,1) | 5355.001 | 2 | 162 | (0,0,0,1,2) | 0 | | | | | |
| 53 | (2,2,2,2,0) | 8925.000 | 2 | 108 | (0,0,0,2,1) | 0 | | 163 | (1,0,0,1,2) | 13.500 | 1 | | | | |
| 54 | (0,0,0,0,1) | 0 | | 109 | (1,0,0,2,1) | 5.625 | 2 | 164 | (2,0,0,1,2) | 47.250 | 1 | | | | |
| 55 | (1,0,0,0,1) | 0 | | 110 | (2,0,0,2,1) | 19.688 | 2 | 165 | (0,1,0,1,2) | 13.500 | 2 | | | | |

**TABLE 6.** ALL-LEVELS RELIABILITY OF FIG. 1.

| $d$ | $r_d$ | $R_d$ |
|---|---|---|
| 1 | 0.041595189 | 0.992447265 |
| 2 | 0.136296380 | 0.950852076 |
| 3 | 0.384884446 | 0.814555696 |
| 4 | 0.429671250 | 0.429671250 |

From $R_d$ in Table 5, thus, we have $R_1$ = 0.992447265, $R_2$ = 0.950852076, $R_3$ = 0.814555696, and $R_4$ =

0.429671250. Note that if any other algorithm is utilized to solve this problem, $r_2$, $r_3$, and $r_4$ will be repeatedly counted two, three, and four times, respectively.

## V. Numerical Experiments

A complete experiment was conducted on 12 benchmark problems, as shown in Fig. 6 to reveal the performance of the proposed algorithm in determining the all-level reliability simultaneously, that is, calculate $R_1$, $R_2$, …, $R_{d\text{MAX}}$ at the same time.

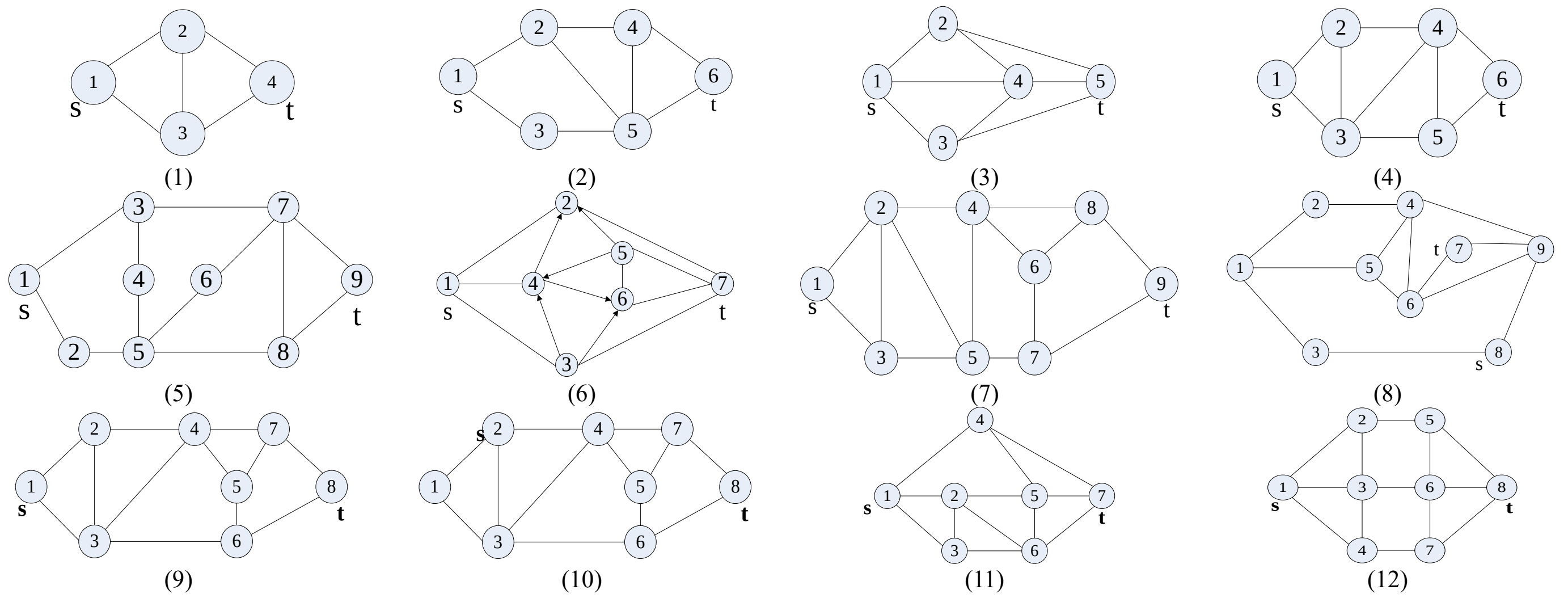

**Figure 6.** 12 binary-state benchmark networks.

### A. Computational Environment

These 12 popular benchmark networks were originally binary-state networks [42]. We redesigned their state distributions to change them to MFNs such that each edge is undirected and multistate. Moreover, without loss generality, each edge is labeled such that the label of edges $e_{i,j}$ for all $i < j$ is less than that of $e_{k,j}$ and $e_{i,l}$ for all $k > (i+1)$, $k < j$, and $j < l$. For example, in Fig. 1, $e_{1,2} = a_1$, $e_{1,3} = a_2$, $e_{2,3} = a_3$, $e_{2,4} = a_4$, and $e_{3,4} = a_5$. Further, $e_{i,j} = e_{j,i}$, for all the edges.

In addition, to ensure a good, time-efficient demonstration with guaranteed results, $\mathbf{D}_{\max}$ is set to 4, the occurrence probability of each state for each vertex is 1/(4+1) = 0.2, and the runtime limit is 60 min, that is, any algorithm is forced to stop if it cannot solve the related benchmark network within 60 min.

Note that there is no existing direct algorithm that can be used to find $R_1$, $R_2$, …, $R_{d\text{MAX}}$ simultaneously.

The proposed all-level direct BAT-based algorithm was programmed in DEV C++ 5.11, executed on 64-bit Windows 10 under Intel Core i7-6650U CPU @ 2.20 2.21 GHz notebook with 16 GB RAM.

The experimental results obtained from the proposed algorithm are listed in Table 7. In Table 7, the notations ID, N, $N_{FC}$, T, and $r_d$ are the benchmark ID, total number of vectors generated from the proposed BAT without using $X_{FC}$, the total number of vectors obtained after using the proposed concept $X_{FC}$, the runtime, and the related reliability, respectively.

Note that $r_d$ is the probability that the exact $d$ units of flow can be sent from nodes 1 to $n$, and

$$R_d = r_d + r_{(d+1)} + \ldots + r_{d\text{Max}} \text{ for all } d \text{ [18].}$$

**TABLE 7.** EXAMPLE THE EXPERIMENTAL RESULTS.

| ID | $n$ | $m$ | $d_{MAX}$ | $X_{FC}$ | T | N | $N_{FC}$ | $r_1$ | $r_2$ | $r_3$ | $r_4$ |
|---|---|---|---|---|---|---|---|---|---|---|---|
| 1 | 4 | 5 | 8 | (1,0,0,1,0) | 0.001 | 3125 | 2999 | 0.1248 | 0.1664 | 0.18496 | 0.192 |
| 2 | 6 | 8 | 4 | (1,0,1,0,0,0,1,0) | 0.063 | 390625 | 374974 | 0.52867328 | 0.236736 | 0.07538944 | 0.02212608 |
| 3 | 5 | 8 | 8 | (1,0,0,0,1,0,0,0) | 0.083 | 390625 | 389999 | 0.10555904 | 0.15064832 | 0.17143552 | 0.18415104 |
| 4 | 6 | 9 | 4 | (1,0,0,1,0,0,0,1,0) | 0.374 | 1953125 | 1874874 | 0.630804992 | 0.21083392 | 0.04730624 | 0.008549888 |
| 5 | 9 | 12 | 4 | (0,1,0,0,1,0,0,0,0,0,1,0) | 70.196 | 244140625 | 234374370 | 0.677284082 | 0.146338673 | 0.021856219 | 0.002871136 |
| 6 | 7 | 14 | 8 | (1,0,0,0,0,1,0,0,0,0,0,0,0,0) | 1859.655 | 6103515625 | 1808545203 | 0.180310919 | 0.184298757 | 0.17555186 | 0.173475457 |
| 7 | 9 | 14 | 4 | (1,0,0,0,1,0,0,0,0,1,0,0,1,0) | 1023.388 | 6103515625 | 1562453953 | 0.722877907 | 0.136000989 | 0.016781315 | 0.00210555 |
| 8 | 9 | 13 | 4 | (1,0,1,1,0,1,0,1,0,0,0,1,0) | 226.574 | 1220703125 | 1171793599 | 0.489854926 | 0.137824511 | 0.084525353 | 0.080937423 |
| 9 | 8 | 12 | 4 | (0,1,0,0,0,1,0,0,0,0,1,0) | 38.862 | 244140625 | 234371870 | 0.685703176 | 0.164227637 | 0.027893613 | 0.005745328 |
| 10 | 8 | 14 | 4 | (0,1,0,0,0,1,0,0,0,1,0,0,1,0) | 1050.259 | 6103515625 | 1562451449 | 0.736874789 | 0.131009741 | 0.013628981 | 0.000885007 |
| 11 | 7 | 12 | 8 | (1,0,0,0,1,0,0,0,0,0,0,0) | 43.278 | 244140625 | 244139999 | 0.222325867 | 0.198123061 | 0.17530241 | 0.167785824 |
| 12 | 8 | 13 | 4 | (1,0,0,0,1,0,0,0,0,1,0,0,0) | 209.367 | 1220703125 | 1218749374 | 0.595187932 | 0.234730638 | 0.085938455 | 0.036154474 |

## B. *Reliability*

**TABLE 8.** EXAMPLE EXPLAINS HOW THE DIFFERENCES NETWORK STRUCTURES AFFECT NETWORK RELIABILITY.

| $n$ | Parallel | Series |
|---|---|---|
| 1 | 0.9 | 0.9 |
| 2 | 0.99 | 0.81 |
| 3 | 0.999 | 0.729 |
| 4 | 0.9999 | 0.6561 |
| 5 | 0.99999 | 0.59049 |
| 6 | 0.999999 | 0.531441 |

From Table 7, we can derive that $0 < R_i < R_j$ if $i > j$. The reason is that $R_j - R_i = r_j + r_{(j+1)} + \ldots + r_{(i+1)}$.

Also, more edges do not guarantee larger reliability. For example, assume that each edge has a reliability of 0.9. Then, we considering Table 8 where $n$ = 1, 2, …, 6, and the second and third columns are the related reliabilities that $n$ edges are connected in parallel and series, respectively.

Hence, the network structure, not the number of edges, determines the network reliability, and such observations are helpful in designing reliable network structures.

If we take a close look, we can find that the ratio $n/m$ increases the reliability because the related network tends to be a complete graph, that is, less sparsity is more reliable.

*C. Runtime*

From Table 7, because the MFN reliability is NP-hard and # P-hard, we can observe that the more edges the more running time, which increases exponentially with the number of edges. Trivially, the runtime started to increase from $m$ = 14.

From the running time T presented in Table 7, the proposed all-level direct BAT almost can obtain $R_1$, $R_2$, …, $R_{d\text{MAX}}$ simultaneously within 30 min, except Fig. (6). Because there are at most $5^{14}$ = 6,103,515,625 vectors in Fig. (6) obtained from the proposed BAT for $m$ = 14, the number of vectors would be too large for any other exact-solution algorithms in the slim possibility that there is one.

The proposed algorithm can also handle 2,198,145.961 vectors (generated from the proposed BAT) per second on average. Hence, from the runtime and the number of benchmark networks, it can be ascertained that the proposed BAT can have all-level reliability.

The primary reason for this is that the proposed direct algorithm can solve the problems directly without requiring the use of IET or SDP, both of which are NP-hard. In contrast, the traditional algorithms for $d$-MP for all $d$ can include $P_1$, $P_2$, …, $P_{d\text{MAX}}$ in the row, which necessitate an implementation of the the IET or SDP to obtain $R_1$, $R_2$, …, $R_{d\text{MAX}}$ sequentially. Moreover, as mentioned in Section IV, all other algorithms take $k$ time to calculate $r_\text{k}$ for $R_1$, $R_2$, …, $R_{d\text{MAX}}$ for $k$ =1, 2, …, $d_\text{MAX}$.

## VI. Conclusions

Networks are indispensable to daily life. The reliability of level $d$, that is, $R_d$, is the probability that the network can send at least $d$ units of flow from the source to the sink node. Hence, reliability plays an important role in evaluating, managing, and designing the performance of all types of network applications. This paper presents a novel all-level reliability problem to calculate $R_1$, $R_2$, …, $R_{d\text{MAX}}$ simultaneously to help the decision maker determine the best value possible for $d$.

To solve this novel problem, a new all-level direct multistate BAT algorithm was proposed. Based on this, the maximum-flow algorithm, and first connected vector, the proposed algorithm can solve the proposed problem with time complexity $O(\varpi n\log n)$, where $\varpi << m^{d\text{MAX}+1}$ is the number of possible vectors obtained from the proposed algorithm and $O(n\log n)$ is the time complexity of the maximum-flow algorithm.

From the numerical experiment conducted on 12 benchmark networks, the proposed all-level direct multistate BAT can calculate the all-level reliability simultaneously within 10 h. In addition, these outcomes also confirm that the proposed algorithm spent only 30 min to deal with one vector generated from the BAT, that is, the proposed algorithm has the potential to solve larger-size problems if the number of disconnected vectors can be reduced and feasible vectors can be verified.

In future work, the proposed all-level direct multistate BAT will be further improved and extended from MFN to multi-commodity MFN (MMFN) and multi-distribution MMFN, with directed edges and special conditions such as budget, speed, time, heterogeneous edges, and unreliable vertices.